\documentclass{article}
\pdfoutput=1
\usepackage{amssymb}	%
\usepackage{latexsym}	%
\usepackage{amsmath}

\usepackage{microtype}  %

\usepackage{hyperref}   %

\usepackage{array}
\newcolumntype{M}[1]{>{\centering\arraybackslash}m{#1}}
\newcolumntype{N}{@{}m{0pt}@{}}

\usepackage[backend=biber,natbib=true,style=authoryear-comp,citestyle=authoryear]{biblatex}
\DeclareSourcemap{
  \maps[datatype=bibtex]{
    \map[overwrite=true]{
      \step[fieldset=address, null]
      \step[fieldset=language, null]
    }
  }
}

\usepackage[utf8]{inputenc} %
\usepackage{csquotes}
\usepackage[english]{babel}

\usepackage{graphicx}

\bibliography{../../FPU}  %

\title{
Formalizing Preference Utilitarianism in Physical World Models\footnote{The final publication in the journal \textit{Synthese} is openly accessible at Springer via \url{http://dx.doi.org/10.1007/s11229-015-0883-1}. This version contains insignificant changes.}}
\author {Caspar Oesterheld\\\texttt{caspar.oesterheld@uni-bremen.de}}
\date{\today}

\begin{document}

\maketitle    %

\begin{abstract}
Most ethical work is done at a low level of formality. %
This makes practical moral questions inaccessible to formal and natural sciences and can lead to misunderstandings in ethical discussion.
In this paper, we use Bayesian inference to introduce a formalization of preference utilitarianism in physical world models, specifically cellular automata.
Even though our formalization is not immediately applicable, it is a first step in providing ethics and ultimately the question of how to ``make the world better'' with a formal basis.%
\end{abstract}


\section{Introduction}

Usually, ethical imperatives are not formulated with sufficient precision to study them and their realization mathematically. \parencites[297]{McLaren2011}[251]{Gips2011}
In particular, it is impossible to implement them on an intelligent machine to make it behave benevolently in our universe, which is the subject of a field known as \textit{Friendly AI} \parencites[e.g. see][2]{Yudkowsky2001} or \textit{machine ethics} \parencites[e.g. see][1]{Andersons2011}.
Whereas existing formalizations of utilitarian ethics have been successfully applied to economics, they are incomplete due to the nature of their dualistic world model in which agents are assumed to be ontologically fundamental.%

In this paper however, we take the following steps towards a workable and simple formalization of preference utilitarianism\footnote{For introductions to and ethical discussions of the underlying notion of preference utilitarianism see \citet{TomasikVideoGameChars2015,TomasikHedVsPrefUtil2015}.
} in physical world models:

\begin{itemize}
\item We describe the problem of informality in ethics and the shortcomings of previous dualist approaches to formalizing utilitarian ethics (section \ref{sec:problem}).
\item We justify cellular automata as a world model, use Bayes' theorem to extract utility functions from a given space-time embedded agent and introduce a formalization of preference utilitarianism (section \ref{sec:idea}).
\item We compare our approach with existing work in ethics, game theory and artificial intelligence (section \ref{sec:relWork}). Our formalization is novel but nevertheless relates to a growing movement to treat agents as embedded into the environment.
\end{itemize}

\section{The problem of formalizing ethics in physical systems} %
\label{sec:problem}

Discussion on informally specified moral imperatives can be difficult due to different interpretations of the texts describing the imperative. Thus, formalizing moral imperatives could augment informal ethical discussion. \parencites[251]{Gips2011}{Anderson2011}{Dennett2006}[19]{Moor2011}

Furthermore, science and engineering answer formally described questions and solve well-specified tasks, but are not immediately applicable to the informal question of how to make the world ``better''.

This problem has been identified in economics and game theory, which has led to some very useful formalizations of utilitarianism \parencites[e.g.][]{Harsanyi1982}. %

\begin{figure}
\begin{center}
\includegraphics[width=0.8\linewidth]{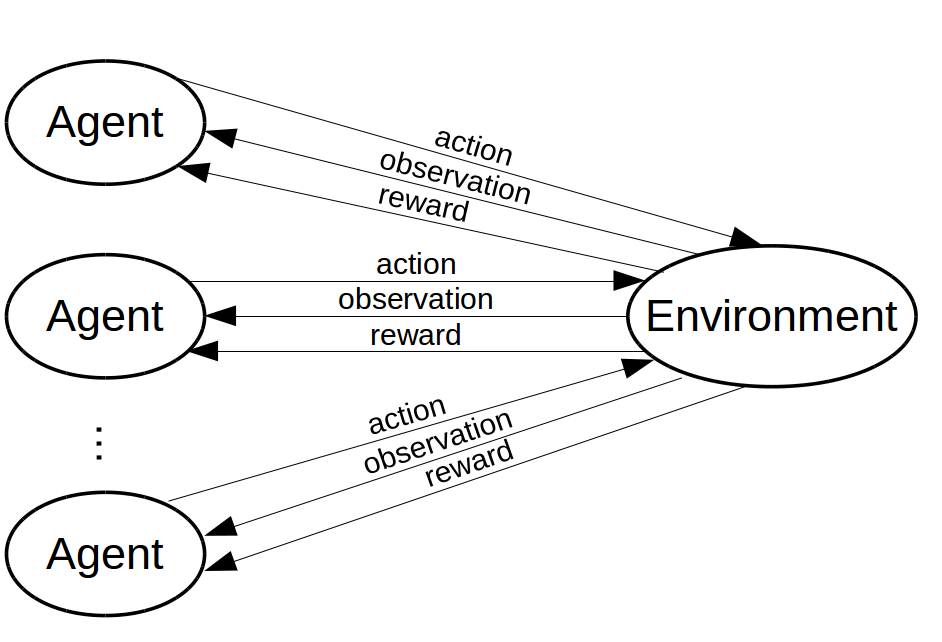}
\end{center} 
\caption{The classic agent-environment-model}
\label{fig:AgentEnvironment}
\end{figure}

However, their formalization relies on consciousness-matter dualism: The agents are not part of the physical world or embedded into it, so that their thoughts or computations can not be influenced by physical laws. Also, agents' utility functions are assumed to not depend on the agents (or their physical configurations) themselves. These are typical assumptions in game theory. After all, game theory is about games, in which players are not actually inside the game, nor can they decide themselves what goals to pursue. This classic \mbox{(multi-)agent-environment} model is depicted in figure \ref{fig:AgentEnvironment}.

Our world, however, is (usually presumed to be) a purely physical system: %
Ethically relevant entities (animals etc.) are embedded in the environment. For example, our brains behave according to the same laws of physics as the rest of the world. Also, happiness and preferences are not given by predetermined utility functions or rewards from the environment, but are the result of physical processes in our bodies. Therefore, dualist descriptions and formalizations leave questions unanswered: \parencite[compare][]{Orseau2012}

\begin{itemize}
\item What objects are ethically relevant? (What are the agents of our non-dualist world?)
\item What is a space-time embedded agent's or, more generally, an object's utility function?
\end{itemize}

Thus, even though classic formalizations of (preferentist) utilitarianism in the agent-environment-model can formalize the vague notions of goals and preferences with utility functions, these formalizations are incomplete, at least in our physical, non-dualist world.

\section{A Bayesian approach to formalizing preference utilitarianism in physical systems}%
\label{sec:idea}

\subsection{Cellular automata as non-dualist world models}

To overcome the described problems of dualist approaches to utilitarianism, we first have to choose a new, physical setting for our ethical imperative. Instead of employing string theory and other contemporary theoretical frameworks, we choose a model that is much more simple to handle formally: cellular automata. These have sometimes even been pointed out to be candidates for modeling our own universe, \parencites[][ch. 9]{NKS}{Schmidhuber1999}{Zuse1967}{Zuse1969} but even if physics will prove cellular automata to be a wrong model, they may still be of instrumental value for the purpose of this paper. \parencites[compare][pp. 70f., 77-79]{Downey2012}[][ch. 8]{Hawking2010}

For detailed introductions to classic cellular automata with neighbor-based rules, see \citet{NKS} or \citet[ch. 7]{natureofcode} for a purely informal and \citet{Wolfram1983} for a slightly more technical treatment that focuses on one-dimensional cellular automata. In section \ref{sec:ca}, we will consider a generalized and relatively simple formalism, which is not limited to rules that only depend on neighbors of a cell.%

In CA, it is immediately clear that for a (preference) utilitarian morality we have to answer the questions that are avoided by assuming a set of agents and their utility functions to be known from the beginning. It also frees us from many ethical intuitions that we build up specifically for our own living situations and reduces moral intuition to its very fundamentals.

\begin{figure}
\begin{center}
\includegraphics[width=0.4\linewidth]{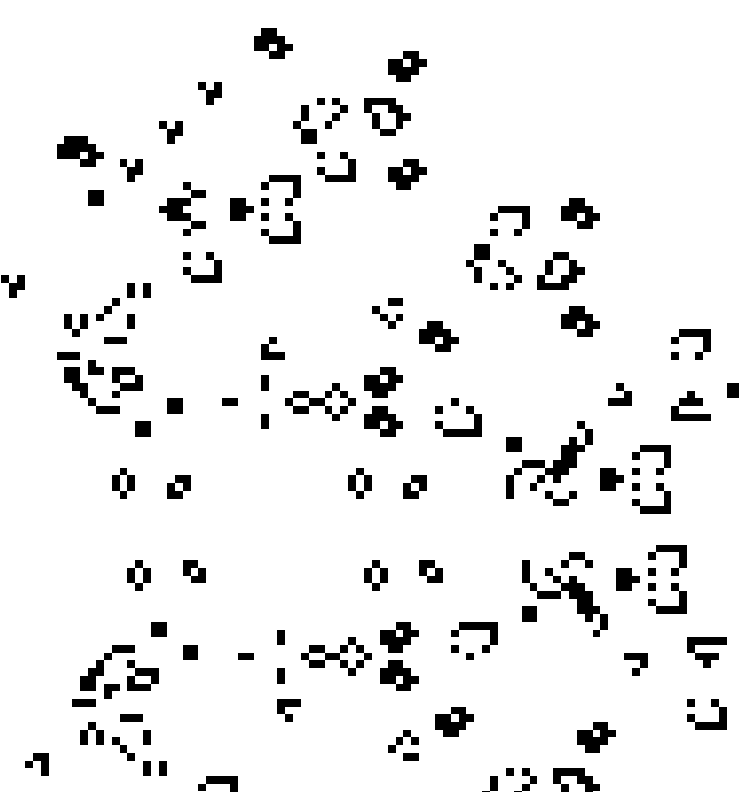}
\end{center}
\caption{A state of a two-dimensional cellular automaton. It is very unclear, what agents are and which preferences they have. Adapted from \url{http://en.wikipedia.org/wiki/Conway\%27s_Game_of_Life\#mediaviewer/File:Conways_game_of_life_breeder.png}}
\label{fig:GOLstate}
\end{figure}

Figure \ref{fig:GOLstate} shows a state of a cellular automaton illustrating the problem of defining utilitarianism or any other ethical imperative in physical models. Clearly, many intuitions are very difficult (if not impossible) to formulate universally and precisely in CA. Thereby, the required formality helps in choosing and defining an ethical imperative.

\subsubsection{A formal introduction to cellular systems}
\label{sec:ca}

We now introduce some very basic notation and terminology of cellular systems, a generalization of classic cellular automata, thus setting the scene for our ethical imperative.

For given sets $A$ and $B$, let $A^B$ denote the set of functions from $B$ to $A$. A \textit{cellular system} is a triple $(C,S,d)$ of a countable set of cells $C$, a finite set of cell states $S$ and a function $d\!: S^C \rightarrow S^C$ that maps a \textit{world state} $s\!:C\rightarrow S$ onto its successor. So basically a world consists of a set of cells that can have different values and a function that models deterministic state-transitions.\footnote{The choice of deterministic systems was made primarily to simplify the formalization. It appears to be unproblematic to transfer formal preference utilitarianism to non-deterministic systems, but defining non-deterministic cellular automata themselves is a little more difficult.}

Cells of cellular systems do not necessarily have to be on a regular grid and computing new states does not have to be done via neighbor-based lookup tables. This makes formalization much easier. %

But before anything else, we have to define structures which represent objects in our cellular systems. A \textit{space} $\mathit{Spc}\subseteq C$ in a cellular system $(C,S,d)$ is a finite subset of the set of cells $C$. A \textit{structure} $\mathit{str}$ on a space $\mathit{Spc}$ is a function $\mathit{str}\! :\mathit{Spc}\rightarrow S$ that maps the cells of the space onto cell values. 

A \textit{history} is a function $h\!:\mathbb{N}\rightarrow S^C$ that maps natural numbers as time steps onto states of the system. For example, the history $h_s$ of an initial state $s$ can then be defined recursively by $h_s(n)=d(h_s(n-1))$ for $n\geq 1$ with the base case $h_s(0)=s$.

\subsection{Posterior probabilities and the priority of a (given) goal to a given agent}

Before extracting preferences from a given structure, we have to decide on a model of preferences. Preferences themselves are mere orderings of alternative outcomes or lotteries over these outcomes with the outcomes being entire histories $h\in \left( S^C \right) ^{\mathbb{N}}$ in our case. The problem is that this makes it difficult to compare two outcomes when the preferences of multiple individuals are involved. To be able to make such comparisons, we move from orderings to utility functions $u:\left( S^C \right)^{\mathbb{N}} \rightarrow \mathbb{R}$ that map histories of the world onto their (cardinal) utilities.\footnote{Other codomains of utility functions seem possible as long as they are subsets of a totally ordered vector space over $\mathbb{R}$. Intervals like $[0,1]$ seem specifically suitable, because they avoid problems of infinite utility and allow for normalization. \parencite{Isbell1959}} This will make it possible to just add up the utilities of different individuals and then compare the sum among outcomes. This by no means ``solves'' the problem of interpersonal comparison of utility. Rather, it makes it more explicit. For example, a given set of preferences is represented equally well by $u$ and $2\cdot u$, but \textit{ceteris paribus} $2\cdot u$ will make the preferences more significant in summation. Different approaches to the problem have been proposed. \parencite{Hammond1989} In this paper we will ignore the problem (or hope that the fair treatment in determining all individuals' utility functions induces moral permissibility). Now we ask the question: Does a particular structure $\mathit{str}$ want to maximize some utility function $u$?

It is fruitful to think about how one would approach such questions in our world, when encountering some very odd organism.
At least one possible approach would be to put it into different situations or environments and see what it does to them.  If the structure increases some potential utility function in different environments, it seems as if this utility function represents an aspect of the structure's preferences.\footnote{Alternatively, one can try to avoid this hypothetical experiment by predicting the organism's behavior. For example, one could try to ask the organism what it would do or infer its typical behavior from its internals. %
} %

However, for some utility functions it is not very special that their values are increased and then it might just be coincidence that the structure in question also does so. For example, it is usually not considered a structure's preference to increase entropy even if entropy increases in environments including this structure, because an increase in entropy is extremely common with or without the structure.

Also, we feel that some utility functions are less likely than others by themselves, e.g. because they are very complex or specific.

But how can we formally capture these notions?

Since uncertainty is involved, we interpret the degree to which a utility function $u$ is important to some structure $\mathit{str}$ that exists at time step $i$ as the posterior probability of that utility function given the structure, a probability we denote by $P(u|\mathit{str}@i)$, where $\mathit{str}@i$ denotes the event that $\mathit{str}$ exists in time step $i$.\footnote{Including $i$ into the data is important, because otherwise identical structures at different points in time would have identical utility functions. This is a problem, when the utility function $u$ is applied to the whole history, because then structures cannot have preferences about themselves (``personal happiness'') without also having preferences about all other identical structures (at the same place). An alternative would be to apply utility functions only to the part of the history from the point of the existence of the structure onwards, so that identical structures at different points in time have equal utility functions that are applied differently. However, it seems like this neglects that the past can depend on the action of an agent in the present, as illustrated in Newcomb's paradox by \citet{Nozick1969}.} Here, the utility function is interpreted as a hypothesis about the structure's ``true intentions''\footnote{\label{ftn:mutEx}Intuitively, some structures can have more than one utility function, while others have no utility function at all. One way to model this would be to understand different utility functions as events in separate sample spaces. So, the sum $\sum_u P(u|\mathit{str})$ could vary among different structures $\mathit{str}$. A similar scenario is the inference of multiple diseases from a set of symptoms. \parencite{Charniak1983} While some individuals may have no diseases or preferences at all, others may be thought of as having more than one disease or utility function. In more technical terms, for each utility function there would be a sample space of having that utility function and not having that utility function.

In this paper however, we will assume mutual exclusivity and collective exhaustiveness of utility functions. All utility functions live in the same sample space and thus $\sum_u P(u|\mathit{str}) =1$ for all structures $\mathit{str}$. This does not mean that all structures have equal moral standing: The idea is that ``meaningless'' structures $\mathit{str}$ have high $P(u|\mathit{str})$ only for constant utility functions $u$, i.e. for ``don't care''-utility functions, which are irrelevant for decision making.
}. In a purely physical, non-dualist world there is nothing but the structure itself, of course. Therefore, the ``true intentions'' do not really exist, which makes it still hard to know what $P(u|\mathit{str}@i)$ is supposed to mean. To finally overcome this problem, we will equate intention and purpose, i.e. we equate the following interpretations of $u$ as a hypothesis explaining the data $\mathit{str}@i$: \parencite[compare][pp. 289ff., 299f., 318, 320f.]{Dennett1989}
\begin{itemize}
\item The utility function $u$ is the goal of structure $\mathit{str}$
\item Maximizing $u$ was the goal of an entity that chose $\mathit{str}$.
\end{itemize}
The second interpretation is more useful, because it describes a data-generating process and thus comes closer to typical statistical models.
Thus, we have to find the posterior probability of some model (a utility function) given some data (a structure). For this problem Bayes' theorem suggests itself, because it provides an equation for posterior probabilities. In our case, Bayes' theorem can be used to infer the likelihood that some utility function was a goal when a structure was chosen from some priors and the likelihood of choosing the structure given that the goal is to maximize the utility function. Specifically, Bayes' theorem gives us
\begin{equation}\label{eq:Bayes}
P(u|\mathit{str}@i) = \frac{P(\mathit{str}@i|u)\cdot P(u)}{P(\mathit{str}@i)},
\end{equation}
where $P(u)$ and $P(\mathit{str}@i)$ are prior probability distributions of utility functions and structures, respectively, and $P(\mathit{str}@i|u)$ is the probability of (some hypothetical entity choosing) $\mathit{str}$ at time step $i$ when $u$ is to be maximized. 
Whereas $P(u|\mathit{str}@i)$ is very hard to grasp intuitively, it is more clear what the probability distributions on the right hand side of the equation mean. Nevertheless, they do not correspond to measurable probability distributions like the results from rolling a dice. Indeed, $P(u)$ and $P(\mathit{str}@i)$ are ultimately \textit{subjective} \parencites[e.g. see][pp. 1f.]{Olshausen2004}[][9]{Robert1994} and $P(\mathit{str}@i|u)$ depends on what exactly the hypothesis $u$ is supposed to express,
thus leaving our ethical imperative parametrized by these distributions. 

Nonetheless, there seem to be canonical approaches. $P(\mathit{str}@i|u)$ should be understood as the probability that $\mathit{str}$ is chosen at time step $i$ by an approximately rational agent that wants to maximize $u$. So, structures that are better at maximizing or more suitable for $u$ should receive higher $P(\mathit{str}@i|u)$ values. This corresponds to the assumption of (approximate) rationality in Dennett's intentional stance. \parencite[][pp.21,49f.]{Dennett1989} Unfortunately, the debate about causal and evidential decision theory \parencite[e.g.][ch. 9]{Peterson2009} shows that formalizing the notion of rational choice is difficult.

The prior of utility functions $P(u)$ on the other hand should denote the ``intrinsic plausibility'' of a goal $u$. That does not have to mean defining and excluding ``evil'' or ``banal'' utility functions. In the preference extraction context, utility functions are models or hypotheses that explain the behavior of a structure. And Solomonoff's formalization of Occam's razor is often cited as a universal prior distribution of hypotheses. \parencite{Legg1997} It assumes complicated hypotheses (utility functions), i.e. ones that require more symbols to be described in some programming language, to be less likely than simpler ones.

If utility functions are conceived of as competing hypotheses (see footnote \ref{ftn:mutEx}), then \[
P(\mathit{str}@i)=\sum_u P(\mathit{str}@i|u) \cdot P(u). \]
Otherwise, $P(\mathit{str}@i)$ could potentially be chosen more freely.

Finally, note how Bayes' theorem catches our intuitions from above, especially when assuming probability distributions similar to the suggested ones: When some structure $\mathit{str}$ maximizes some utility function $u$ very well, then $P(\mathit{str}@i|u)$ and thereby the relevance of the utility function to the object would increase. On the other hand, if many other structures are comparably good, then the probability for each one to be chosen when given the utility function is smaller (due to the sum of the probabilities of all possible structures on a given space and time step being 1) and the probability of the utility function being a real preference would decrease with it. Finally, multiplying by $P(u)$ catches abstruse utility functions, e.g. utility functions that are specifically suited to be fulfilled by the structure in question.

\subsection{An individual structure's welfare function}

Having introduced a way of determining how likely it is that some utility function is the utility function of some object, we define the welfare $U_{\mathit{str}@i}$ of a structure $\mathit{str}$ that exists at some step $i$ of a history $h$, as the weighted sum over all utility functions
\begin{equation} \label{eq:onestrexputil}
U_{\mathit{str}@i}=\sum_u P(u|\mathit{str}@i) u(h),
\end{equation} 
where $h$ is the history and the sum is over all theoretically possible utility functions $u:(S^C)^{\mathbb{N}} \rightarrow  \mathbb{R}$.

We call this term \textit{expected utility}, because this expression is generally used for adding utilities weighted by their likelihoods, which is a common concept. However, the term usually suggests that there is also an \textit{actual utility}. In our case of ascribing preferences to physical objects however, no such thing exists. We only \textit{imagine} there to be some real utility or welfare functions and that we use Bayesian inference to find them. But in fact, the structure itself is all there exists and thus the expected utility is as actual as possible.

The sum in the term for expected utility is over an uncountably infinite set, which can only converge when only countably many summands are non-zero.\footnote{If $\sum_u P(u)=1$ (see footnote \ref{ftn:mutEx}), then this is given automatically. Also, if Solomonoff's prior is chosen for $P(u)$, all incomputable utility functions have zero probability. Since the set of computable functions is countable, only countably many summands could possibly be non-zero.} Some other concerns are described in footnote \ref{ftn:riemannseriestheorem} and addressed in footnote \ref{ftn:SeriesvsValue}.

\subsection{Summing over all agents}

The utilitarian imperative is to maximize a global welfare function that is the sum of all individuals' welfare functions. We already defined the welfare function of single structures. So next we have to define what the set of all agents is and how to sum over it. As foreshadowed before, we will consider all possible structures of a cellular automaton using equation \ref{eq:onestrexputil} and rely on (intuitively) irrelevant ones to receive high $P(u|\mathit{str}@i)$ values only for constant and therefore irrelevant utility functions $u$ (see footnote \ref{ftn:mutEx}).
To sum the utility over all agents, we not only have to sum over all structures in a particular state, but first over all (discrete) time steps of the history of the cellular automaton world and only then over all structures in every state. This way, we sum the welfare of all agents ever coming into existence. For the summands, we can insert the term obtained in equation \ref{eq:onestrexputil}
\begin{equation} \label{eq:FPUmain}
\sum_{i} \sum_{\mathit{str}@i} U_{\mathit{str}@i} = \sum_{i} \sum_{\mathit{str}@i} \sum_u P(u|\mathit{str}@i) u(h),
\end{equation}
where $U_{\mathit{str}@i}$ denotes the welfare or utility of the structure $\mathit{str}$ that exists at time step $i$, the first sum is over all integers functioning as time steps, the second is over all structures in $h(i)$ and the third over all possible utility functions.\footnote{More precisely, but less elegantly, one could write
\[
\sum_{i=0}^\infty ~ \sum_{\mathit{Spc} \in Fin(C) } ~ \sum_{u :( S^C )^{\mathbb{N}} \rightarrow \mathbb{R}} u(h) P(u|(h(i)|_{\mathit{Spc}})@i),
\]
where $Fin(C):=\{ A \subseteq C | |A|\in \mathbb{N} \}$ is the set of finite subsets of $C$ and $h(i)|_{\mathit{Spc}}\! :\mathit{Spc}\rightarrow S:c\mapsto h(i)(c)$ is the restriction of the state $h(i)$ to the space $\mathit{Spc}$ and therefore the structure on that space.
} So, our formalization of the main imperative of preference utilitarianism turns out to be nothing more than maximizing \textit{global, all-time expected utility} (of every space-time embedded agent that ever comes into existence). In general, the value of the series depends on the order of these infinite sums.\footnote{\label{ftn:riemannseriestheorem} Specifically, the Riemann series theorem states that any conditionally convergent series can be reordered to have arbitrary values.} Also, the series can diverge. Nonetheless it may still be usable for comparing histories in many cases.\footnote{\label{ftn:SeriesvsValue}It is very important to differentiate the series from its value. Otherwise, one may identify the series with positive or negative infinity or as being undefined. Two infinite values of the series would then
not be comparable anymore, which \citet{Bostrom2011} identified as a problem for (consequentialist) ethics. But this
problem can sometimes be eliminated by comparing the series itself to another. In this particular case, a history $h$ is better than another history $h'$, if
\[
\sum_i \sum_{\mathit{Spc}} \sum_u u(h) P(u|(h(i)|_{\mathit{Spc}})@i) - u(h') P(u|(h'(i)|_{\mathit{Spc}})@i) >0,
\]
where $h(i)|_{\mathit{Spc}}\! :\mathit{Spc} \rightarrow S: c \mapsto h(i)(c)$ denotes the restriction of $h(i)$ to $\mathit{Spc}$, i.e. the structure on $\mathit{Spc}$ in the state $h(i)$. If no such relation can be established then the two histories are arguably incomparable or may be called approximately equally good. Again, the ordering could be important in some cases, see footnote \ref{ftn:riemannseriestheorem}. 
}%

\section{Related work}
\label{sec:relWork}

Preferentist utilitarianism has become a common form of utilitarianism in the second half of the 20th century, with the best known proponents being Hare and Singer. However, the intuitions underlying the presented formalization are different from the most common ethical intuitions in preference utilitarianism. Since our formal preference utilitarianism is not meant to describe a decision procedure for humans (or, more generally and in Hare's (1981, pp. 44f.) \nocite{Hare1981} terminology, non-``archangels''), we do not consider an application-oriented utilitarianism like Hare's two-level consequentialism. \citep[p. 25ff.]{Hare1981} %
Also, most preference utilitarians ascribe preferences only to humans (or abstract agents) and do not contain prioritization among individuals, \citep[p. 46]{Harsanyi1982} or they use a low number of classes of moral standing. \citep[pp. 101ff., 283f.]{Singer1993} Whereas some have pointed out that a variety of behavior and even trivial systems can be viewed from an ``intentional stance'', \parencites[][]{Dennett1971}[][especially pp. 29f.;]{Dennett1989}[compare][pp. 52ff.]{Hofstadter2007} %
only relatively recent articles in preference utilitarianism have discussed the connection between goal-directed behavior and ethically relevant preferences and with the universality of the former pointed out the potential universality of the latter. \parencites[][ch. 7]{TomasikHedVsPrefUtil2015}[][ch. 4, 6]{TomasikVideoGameChars2015}[][]{TomasikSufferingPhysics2015} This idea is an important step when formalizing preference utilitarianism because otherwise one would have to define moral standing depending on other, usually binary, notions: being alive, the ability to suffer \citep[ch. 17 note 122]{Bentham1823}, personhood \citep[ch. 1]{Gruen2014}, free will, sentience and (self-)consciousness \citep[pp. 101ff.]{Singer1993} or the ability of moral judgment.
However, all of them seem to be very difficult to define (universally) in physical systems in the intended binary sense.\footnote{For example, \citet[pp. 823-825, 1178-1180]{NKS} and \citet{Emmeche1997} discuss the property of life, %
\citet[pp. 9-24, 51-54]{Hofstadter2007} discusses consciousness and \citet[p. 5]{Arneson1998} discusses personhood.} Also, continuous definitions of these terms are often connected with goal-directed behavior. \parencites[][ch. 4]{TomasikVideoGameChars2015}[][p. 1136]{NKS}

Whereas most ethical work is conducted informally, \parencites[][297]{McLaren2011}[][251]{Gips2011} there has been some formal work at the intersection of (utilitarian) ethics, game theory and economics, most notably by \citet{Harsanyi1982}.
Some formalization has also been conducted in the realm of machine ethics. \parencites[][]{Anderson2004}[][pp. 245ff.]{Gips2011}
However, influenced by game theory and dualist traditions in philosophy, they are based on the classic agent-environment-model as displayed in figure \ref{fig:AgentEnvironment} and assume utility functions (or even the utilities in different trajectories themselves) as given by the world model. Nonetheless, there is at least one parallel: all models of utilitarianism contain the notion of summing the utility over all agents. As shown in figure \ref{fig:comparison}, both the definition of \textit{all agents} and how to obtain the utility or welfare of an agent differ among formalizations.

\begin{figure}
\begin{center}
\begin{tabular}{|M{3cm} M{3cm}|N}  %
	\hline
	sum over all agents & utility of an agent& \\[10pt]
	\hline
	${\displaystyle \sum_i \sum_{\mathit{str}@i}}$ & ${\displaystyle \sum_u P(u|\mathit{str}@i) u(h)}$ & \\[20pt]
	${\displaystyle \sum_{\text{agent}~n\in N}}$ & ${\displaystyle U_n}$ & \\[20pt]
	${\displaystyle \sum_{\text{agent}~n\in N}}$ & ${\displaystyle w_n \cdot U_n}$ & \\[20pt]
	\hline
\end{tabular}
\end{center}
\caption{Comparison between formalizations of utilitarianism. The first row shows the formalization of this paper, the second row is adapted from \citet[p. 46]{Harsanyi1982}, and the third row from \citet[p. 245]{Gips2011}. The utility of an agent $n$ is denoted by $U_n$ and its weight by $w_n$.}
\label{fig:comparison}
\end{figure}

In Artificial Intelligence, the idea of \textit{learning} preferences has become more popular, e.g. see \citet{Fuernkranzetal2010} and \citet{Nielsenetal2004} for technical treatments or \citet[pp. 192ff.]{Bostrom2014} for an introduction in the context of making an AI do what the engineers value. However, most of the time, the agent is still presumed to be separated from the environment.

Nonetheless, the idea of evaluating space-time-embedded intelligence is beginning to be established in artificial (general) intelligence, \parencite{Orseau2012} which is closely related to the probability distribution $P(\mathit{str}@i|u)$.

\section{Conclusion} %

By reversing Dennett's intentional stance with Bayes' theorem, we were able to ascribe preferences to physical objects and thus formalize preference utilitarianism in cellular automata. Theoretically, such formalizations can function as a specification for an artificial intelligence or more generally as a basis for ``paradise engineering'' \parencites[e.g. see][124]{Ettinger2009}.
However, there are several potential problems that require further work before such practical applications of our formalization or improved variations of it can be approached:
\begin{itemize}
\item Through sums over all structures, possible utility functions and states and the application of incomputable concepts like Solomonoff's prior in $P(u)$, our formalization is incomputable in theory and practice. So even in simulations of cellular automata our formalization is not immediately applicable. %This makes it difficult to understand to understand its mechanisms.
\item Computing our global welfare function in the real world is even more difficult, because it requires full information about the world on particle level. Also, the formalization must first be translated into the physical laws of our universe. %This could be used to compare ... with intuition.
\item The difficulty to apply our formalization is by no means only relevant to actually using it as a moral imperative. Instead, it is also relevant to discussing our formalization from a normative standpoint: Even though the derivation of our formalization is plausible, it may still differ significantly from intuition. There could be some kind of trivial agents with trivial preferences that dominate comparison of different histories. Because the formalization's incomputability makes it difficult to assess whether such problems are present, further work on its potential flaws is necessary. Based on such discussion, our formalization may be revised or even discarded. In any case, we could learn a lot from its shortcomings especially due to the formalization's simplicity and plausible derivation.
\item We outlined how $P(\mathit{str}@i|u)$ and $P(u)$ could be determined in principle. However, they need to be specified more formally, which in the case of $P(\mathit{str}@i|u)$ seems to require a solution to the problem of normative decision theory. Some problems of our formalization could inspire additional refinements of these distributions.
\end{itemize}

\section*{Acknowledgements}

I am grateful to Brian Tomasik for giving me important comments that led me to
systematize my formalization. I also thank Adrian Hutter for an interesting discussion on the formalization,
as well as Alina Mendt, Duncan Murray, Henry Heinemann, Juliane Kraft and Nils Weller for reading and
commenting on earlier versions of the paper. I owe thanks to the two anonymous reviewers whose comments
and suggestions helped improve and clarify this manuscript.

\newpage
\begin{sloppypar} %
\printbibliography
\end{sloppypar}

\end{document}